\documentclass[10pt]{article}
\usepackage{amssymb}
\usepackage{amsmath}
\usepackage[all]{xy}
\usepackage{amsmath}
\usepackage{amsfonts}
\usepackage{amssymb}
\usepackage{amscd}
\usepackage{amsthm}
\usepackage{latexsym}
\usepackage{amsbsy}
\usepackage{graphicx}
\usepackage{float}
\usepackage[toc,page]{appendix}
\usepackage{caption}
\usepackage{subcaption}

\usepackage{hyperref}
\hypersetup{
	colorlinks,
	citecolor=black,
	filecolor=black,
	linkcolor=black,
	urlcolor=black
}
\makeatletter

\newtheorem{theorem}{Theorem}[section]

\newtheorem{corollary}{Corollary}[theorem]
\newtheorem{lemma}[theorem]{Lemma}

\newtheorem*{remark}{Remark}
\DeclareMathOperator{\tr}{Tr}

\usepackage[most]{tcolorbox}
\newtcolorbox{highlighted}{colback=yellow,coltext=black,breakable}

\usepackage{soul}

\def\section{\@startsection{section}{1}{\z@}{-3.25ex plus -1ex minus
		-.2ex}{1.5ex plus .2ex}{\normalfont\bfseries}}
\def\subsection{\@startsection{subsection}{1}{\z@}{-3.25ex plus -1ex
		minus -.2ex}{1.5ex plus .2ex}{\normalfont\itshape}}
\date{}
\makeatother
\title{Double scaling limits of Dirac ensembles and Liouville quantum gravity}
\author{Hamed Hessam, Masoud Khalkhali, and  Nathan Pagliaroli\\
	Department of Mathematics,  Western University\\
	London, Ontario, Canada\footnote{\emph{Email addresses}:  hhessam@uwo.ca, masoud@uwo.ca, npagliar@uwo.ca. Corresponding author: masoud@uwo.ca.}}

\begin{document}
	\maketitle
	\begin{abstract}
		In this paper we study ensembles of finite real spectral triples equipped with a path integral over the space of possible Dirac operators. In the noncommutative geometric setting of spectral triples, Dirac operators take the center stage as a replacement for a metric on a manifold. Thus, this path integral serves as a noncommutative analogue of integration over metrics, a key feature of a theory of quantum gravity. From these integrals in the so-called double scaling limit we derive critical exponents of minimal models from Liouville conformal field theory coupled with gravity. Additionally, the asymptotics of the partition function of these models satisfy differential equations such as Painlev\'e I, as a reduction of the KDV hierarchy, which is predicted by conformal field theory. This is all proven using well-established and rigorous techniques from random matrix theory.  
	\end{abstract}
	
	\section{Introduction}
	Attempts to construct theories of Euclidean quantum gravity typically involve a partition function where the integration  is over possible topologies or metrics and matter fields. However, these integrals are nonrenormalizable. Some approaches often used in physics involve discretizing the space, metrics, or topologies. Discrete approximations of physical theories have often found much success, such as in lattice gauge theory \cite{lattice gauge theory}. An alternative approach  comes through noncommutative geometry. One may approximate commutative spaces by replacing the algebra of commutative functions on that space by a corresponding algebra of matrices. This was first done with the fuzzy sphere 	
	\cite{fuzzy sphere}. Such constructions exist for other spaces; for an example see the construction of the complex projective plane in \cite{Grosse}.
	
	As mentioned earlier, one would like to construct a path integral over all metrics (and eventually over matter fields), but when a space is ``fuzzified"  its metric loses its meaning. Alternatively, in the framework of spectral triples the role of a metric is played by the Dirac operator, as in Conne's distance formula \cite{Connes-2013}. Barrett first suggested in \cite{Barrett2015} that a toy model for finite noncommutative quantum gravity could be constructed as a well-defined  matrix integral over an appropriate space of   Dirac operators. We refer to these models as \textit{Dirac ensembles}. The goal was that in some appropriate limit Dirac ensembles might connect to some understood physics, thus validating this random fuzzy approximation. The resulting partition functions of these models are matrix integrals, allowing one to use techniques from random matrix theory. In this paper we find that certain Dirac ensembles are dual to minimal models from conformal field theory coupled to gravity in the double scaling limit.

	In random matrix theory the expectation values of observables can  typically be written as a formal summation organized by genus, called the genus expansion, which was first discovered by t' Hooft \cite{t' Hooft}. Taking the large $N$ limit  of matrix models is equivalent to counting various types of planar maps \cite{BIPZ}. As proven in \cite{Second paper}, Dirac ensembles of any dimension have a genus expansion. In particular, the leading order term, found by a large $N$ limit of appropriately scaled quantities, amounts to counting strictly planar maps. In the late 80's and 90's physicists found artifacts of conformal field theory in the large  $N$ limit of matrix models  when coupling constants were tuned to specific critical values \cite{LQG 1,LQG 2,Kazakov multi}. These critical values occur when the genus expansion terms of the log of the partition function fail to be smooth. This is analogous to how critical values are found in statistical mechanics.
	\begin{figure}[htp]
		\centering
		\includegraphics[width=0.3\textwidth,height=100mm]{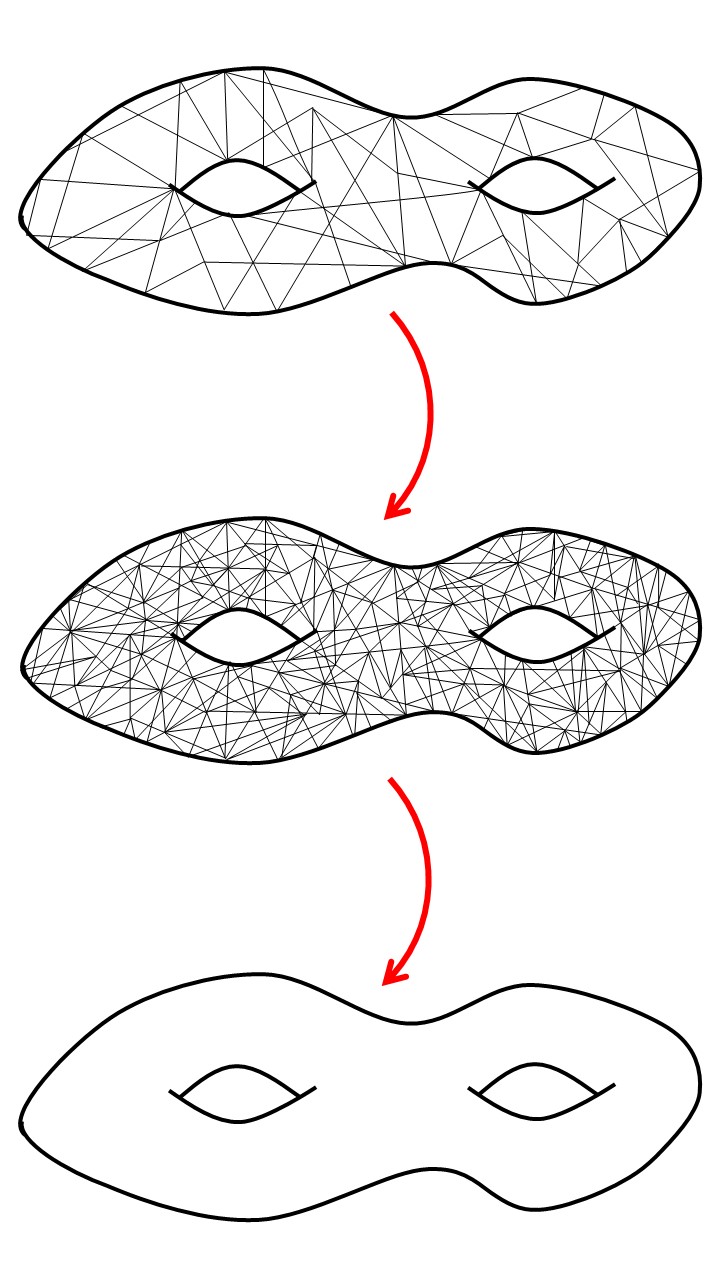}
		\caption{Intuitively if one fine tunes coupling constants of matrix models such that the number of polygons in maps goes to infinity, maps are replaced by smooth surfaces.}
	\end{figure}
	
	We give some intuition for this connection. Formal random matrix models are formal summations of Gaussian integrals, which via Feynman graph techniques can be realized as  the weighted generating functions of maps \cite{BPZ}. A map is a type of embedding of a graph onto a two dimensional  surface. Maps can be seen as a method of discretizing surfaces, and when the number of polygons that make up the map becomes very large, one would expect that one should be able to find a connection to partition functions that sum over   surfaces. The number of polygons in maps can in fact grow by tuning the coupling constants to critical values \cite{Eynard LQG}. Note that these critical values are not the same as those discussed in \cite{Barrett2016,First paper}.

	The authors wish to emphasize that  none of these connections between quantum gravity and random matrices are new, but our goal is to reframe them as a toy noncommutative theory of quantum gravity. We prove in this paper that single trace matrix ensembles emerge from Dirac ensembles when the coupling constants are tuned appropriately. As we will discuss, this is highly non-obvious because the coupling constants of bi-tracial and single trace terms are shared.  This allows one  to recover random matrix models that are dual to minimal models of conformal field theory, an old and well studied connection in physics \cite{Eynard LQG}.  
	
	In Section 2 we give a brief introduction to Dirac ensembles. The Schwinger-Dyson equations and stuffed maps are developed in Section 3. The resolvent function is computed for some examples of Dirac ensembles in Section 4. Section 5 explains how one uses blobbed topological recursion to compute higher order and genus correlation functions. Section 6 highlights the main result of this paper. In this section the critical points  of the examples in Section 5 are discussed as well as the connection to Liouville conformal field theory. In Section 7 we summarize these results and future projects.

	\tableofcontents
	
	\section{Dirac ensembles}
	
	A Dirac ensemble is defined by a fixed finite dimensional real  spectral triple, in which the Dirac operator is allowed to be randomly selected subject to some consistency conditions. Let us briefly recall that a finite real spectral triple is a quintuple  $(\mathcal{A},\mathcal{H},D,\Gamma,J)$ where $\mathcal{A}$ is a finite dimensional complex involutive algebra acting on the Hilbert space $\mathcal{H}$. The Dirac operator is a self-adjoint operator from $\mathcal{H} \rightarrow \mathcal{H}$. The two extra operators $\Gamma$ and $J$, known as the charge conjugation  and the chirality operator play no active role in our analysis so will not be discussed here. For a more detailed explanation see \cite{QFTNCG,Barrett2015,VS, Marcolli}.
	
	The partition function of a Dirac ensemble is given by Barrett and Glaser \cite{Barrett2016} as
	
	\begin{equation}\label{partition function}
		Z = \int_{\mathcal{D}}e^{-\tr V(D)} d D,
	\end{equation}
	where $\mathcal{D}$ is the space of possible Dirac operators, which form a real vector space, and $dD$ is the Lebesgue measure on this space. In general, the Dirac operator can be expressed in terms of gamma matrices tensored with commutators and anti-commutators of Hermitian and skew Hermitian matrices. More precisely, in \cite{Barrett2016} it was found that for any fuzzy spectral triple, the Dirac operator is of the form
	\begin{equation*}
		D = \sum \gamma^{I}\otimes \{K_{I},\cdot\}_{e_{I}}
	\end{equation*}
	where the sum is over increasingly ordered multi-indices, and the following rules apply:
	\begin{itemize}
		\item  If $\gamma^{I}$ is Hermitian,  $\{K_{I},\cdot\}_{e_{I}} = \{H_{I},\cdot \}$, where $H_{I}$ is some Hermitian matrix and $\{\cdot\}$ is the anti-commutator.
		\item If $\gamma^{I}$ is skew-Hermitian,  $\{K_{I},\cdot\}_{e_{I}} = [L_{I},\cdot]$, where $L_{I}$ is some skew-Hermitian matrix and $[\cdot]$ is the commutator.
	\end{itemize}

	Note that the  $H_{I}$ and $L_{I}$ are free variables. As a result, integral (\ref{partition function}) can be expressed  as an integral over the Cartesian  product of the spaces of $N\times N$ Hermitian matrices, $\mathcal{H}_{N}$, and skew-Hermitian matrices $\mathcal{L}_{N}$:
	\begin{equation*}
		Z = \int e^{-\tr V(D)} d D,
	\end{equation*}
	where $dD$ is the Lebesgue measure on the product space of finitely many spaces of Hermitian and spaces of skew-Hermitian matrices. In particular when $p+q$ is equal to one or two we have 
	\begin{equation*}
		dD = \prod_{\ell,r = 1}^{p,q}dH_{\ell}dL_{r}= \prod_{\ell,r = 1}^{p,q} \prod_{i=1}^{N}dH_{r_{ii}}d \text{Im}(L_{\ell_{ii}}) \prod_{i<j}d \,\text{Re}(H_{\ell_{ij}})d\, \text{Im}(H_{\ell_{ij}})\,d \,\text{Re}(L_{r_{ij}})d\, \text{Im}(L_{r_{ij}}).
	\end{equation*}
	The choice of $S(D)$ is left open. However, we are particularly interested in cases where  $S$ is  a polynomial in $D$.\footnote{Note that the use of $(p,q)$ for the KO-dimension of fuzzy spectral triples has no relationship with the integers $(p,q)$ that are used in Kac's table for minimal models in conformal field theory.}
	
	A skew Hermitian matrix can be written as a Hermitian matrix multiplied by the complex unit. Furthermore, since $S(D)$ is a polynomial, and thus the potential is a trace polynomial of the Hermitian and skew-Hermitian matrices seen in $D$, any integral of the above form can be written strictly as a Hermitian multi-matrix integral.  These objects are interesting purely from a random matrix perspective, of which very little is known in general. Relatively recently  some universal properties  were established in \cite{Second paper}. 
	
	Unlike the usual matrix model, in addition to the spectra of the $H$'s we have the spectrum of $D$ to study. As one might guess, there is a deep and not well understood relationship between them. The spectrum of the Dirac operator itself is not fully understood but  displays some universal behaviors as seen in \cite{Second paper} and spectral phase transitions as seen in \cite{Barrett2016,First paper,glaser}. We are interested only in the simplest cases for now, partly because of the lack of analytical tools needed to study multi-matrix models.

	We will consider a one dimensional Dirac ensemble of type $(1,0)$. We emphasize that one could also work with $(0,1)$ just as easily, such as in \cite{First paper}. The type $(1,0)$ signifies that the associated Clifford module of the fuzzy spectral triple has a signature of one.  Such a Dirac ensemble consist of a finite real spectral triple of the form $(M_{N}(\mathbb{C}),M_{N}(\mathbb{C}),D)$ where 
	\begin{equation*}
		D = H\otimes I + I \otimes H,
	\end{equation*}
	and $H$ is some Hermitian $N\times N$ matrix sampled from the probability distribution
	\begin{equation*}
		e^{-\tr V(D)} dH.
	\end{equation*}
	The function $S(D)$ is some polynomial in $D$ and 
	\begin{equation*}
		dH = \prod_{i=1}^{N}dH_{ii} \prod_{i<j}d \,\text{Re}(H_{ij})d\, \text{Im}(H_{ij}).
	\end{equation*}
	
	It is not hard to see that trace powers of $D$ can be written as follows
	\begin{equation}\label{Dirac moments}
		\tr D^{\ell} = \sum_{k=0}^{\ell} {\ell \choose k} \tr H^{\ell -k}\tr H^{k}.	
	\end{equation}
	Thus, the integral 
	\begin{equation*}
		Z = \int_{\mathcal{H}_{N}} e^{-\tr V(D)} dH
	\end{equation*}
	is not just a matrix integral, but more specifically a bi-tracial matrix integral. We will discuss this further in the next section.
	
	We define a  formal matrix integral as the formal summation formed by power series expanding all non-Gaussian terms in the integrand and swapping the order of integration and summation. Such formal sums of Gaussian integrals can be evaluated termwise using Wick's theorem \cite{Zvonkin and Lando,Eynard2018}. Graphically, the coefficients of this formal sum can be realized as a weighted generating function counting stuffed maps \cite{blobbed1}. 
	
	We can define the matrix moments  and higher moments of this ensemble as 
	\begin{equation*}
		\mathcal{T}_{\ell} := \langle \frac{1}{N}\tr H^{\ell}\rangle =\frac{1}{N}\frac{1}{Z} \int_{\mathcal{H}_{N}}\tr H^{\ell} \,e^{-\tr V(D)} dH
	\end{equation*}
	and 
	\begin{equation*}
		\langle \frac{1}{N^{n}}\tr H^{\ell_{1}}\tr H^{\ell_{2}}...\tr H^{\ell_{n}}\rangle :=\frac{1}{N^{n}}\frac{1}{Z} \int_{\mathcal{H}_{N}}\tr H^{\ell_{1}}\tr H^{\ell_{2}}...\tr H^{\ell_{n}}\, e^{-\tr V(D)} dH.
	\end{equation*}
	
	One can further define joint cumulants of higher moments using the classical moment-cumulant  relations. For details see chapter one of \cite{Eynard2018}. These cumulants are denoted $\langle \frac{1}{N^{n}}\tr H^{\ell_{1}}\tr H^{\ell_{2}}...\tr H^{\ell_{n}}\rangle_{c}$ and are the generating functions of connected maps in the sense that the embedded graph is connected.

	The Dirac moments can be computed from matrix moments in this case  via formula (\ref{Dirac moments}). Higher order moments (which were not defined above) can be obtained in the usual manner from higher order cumulants. As proven in \cite{AK}, the matrix moments and cumulants have a genus expansion, i.e.
	
	\begin{equation*}
		\mathcal{T}_{\ell_{1},...,\ell_{q}}= \sum_{g=0}^{\infty}\left(\frac{N}{t}\right)^{2-2g-q}\mathcal{T}^{g}_{\ell_{1},...,\ell_{q}}
	\end{equation*}
	where $t$  is a continuous formal parameter strictly greater than zero. The terms of the genus expansion can be put into generating functions of the form
	\begin{equation*}
		W^{g}_{k}(x_{1},x_{2},...x_{k}) = \sum_{\ell_{1},\ell_{2},...,\ell_{k}=0}^{\infty} \frac{\mathcal{T}^{g}_{\ell_{1},\ell_{2},...\ell_{k}}}{x_{1}^{\ell_{1}+1} x_{1}^{\ell_{2}+1}... x_{1}^{\ell_{1}+1}}.
	\end{equation*}
	
	The terms of this so-called genus expansion can be computed recursively using a process called blobbed topological recursion \cite{AK}. Blobbed topological recursion is a generalization of the similar process of topological recursion \cite{TR} which has gained much interest in the last two decades. For a review we refer the reader to \cite{Borot review}.
	
	In Section \ref{SDE's}, we will derive a set of recursive equations that can relate cumulants and moments (as well as their generating functions) called the Schwinger-Dyson equations. It is a well-known practice in random matrix literature to compute $W_{1}^{0}$ using resolvent techniques. The beauty of (blobbed) topological recursion is that given $W_{1}^{0}$ and $W_{2}^{0}$ (which is often in some sense universal), one can compute any $W_{k}^{g}$ recursively by decreasing Euler characteristic  $-\chi = 2g-2+n$. 
	
	For single trace matrix models this process is well studied and formalized, see \cite{Eynard2018}. However, for multi-trace matrix models we are unaware of any similar reference. In this paper we will explicitly show how to compute $W_{1}^{0}$   and show that $W_{2}^{0}$ has a universal form for bi-tracial matrix models.

	\section{Bi-tracial matrix integrals}
	In this section we will set the groundwork for analyzing bi-tracial matrix models. 
	
	\subsection{Stuffed maps}
	\label{Stuffed maps}
	In this paper we are strictly interested in bi-tracial matrix models since they are the ones of interest for Dirac ensembles. However, one would expect that this analysis can be  extended to higher trace multiplicity. Consider  the following matrix integral over the space of Hermitian matrices
	
	\begin{equation}\label{multi trace ensemble}
		Z =\int_{\mathcal{H}_{N}} e^{-\tr \tilde{V}(H)} dH, 
	\end{equation}
	where the trace of the  potential is a bi-tracial polynomial in $H$
	
	\begin{equation*}
		\tr \tilde{V}(H) = \frac{N \,t_{2}}{2t}\tr H^{2} + \sum_{i=3}^{d} \frac{N}{t} \frac{t_{i}}{i}\tr H^{i} + \sum_{i,j=1}^{d} \frac{t_{i,j}}{ij}\tr H^{i} \tr H^{j}
	\end{equation*}
	where the $t_{i}$'s and $t_{i,j}$'s are coupling constants such that $t_{i,j} = t_{j,i}$. 

		It was first discovered in \cite{BIPZ} that the moments  and cumulants of the random Hermitian matrix ensembles coincide with the generating functions of maps.  The maps of interest for bi-tracial matrix models are called stuffed maps and were first studied in \cite{blobbed1} by Borot, and subsequently in \cite{blobbed2}. Note that stuffed maps are a direct generalization of the maps in \cite{Eynard2018} used for single trace matrix models, thus all the following definitions simplify to the types of maps first considered in \cite{BIPZ}. We now define the building blocks of stuffed maps.

		An elementary 2-cell of topology $(k,h)$ is a connected oriented surface of genus $h$ with $k$ boundaries. For example, a 2-cell with topology $(1,0)$ is a disc. These 2-cells can be ``glued" together by pairing edges of the perimeter to form a surface with an embedded graph. The resulting surface is referred to as a \textit{stuffed map} of topology $(n,g)$ with perimeters $(\ell_{1},..,\ell_{n})$. It is an orientable connected surface with boundaries of lengths  $\ell_{1},..,\ell_{n}$. For more on stuffed maps see \cite{blobbed1}.
		
		We are interested in enumerating these maps. More specifically, we want to count stuffed maps that are glued from  2-cells with the topology of discs and cylinders. This comes from the fact that the Dirac ensembles of interest are bi-tracial matrix models with  appropriate $N$-powers as  coefficients \cite{blobbed1}. We shall refer to these maps as \textit{unstable} stuffed maps. We define $\mathbb{SM}_{k}^{g}(v)$ as the set of all unstable stuffed maps of genus $g$, with $v$ vertices and $k$ boundaries. It was proven in \cite{Second paper} that this set is finite, allowing us to define the following formal series:
		
		\begin{equation}\label{moments}
			\mathcal{T}^{g}_{\ell_{1},...,\ell_{k}} =\sum_{v=1}^{\infty}t^{v}\sum_{\Sigma \in \mathbb{SM}_{k}^{g}(v)}\prod_{i=3}^{d}t_{i}^{n_{i}(\Sigma)} \prod_{i,j=0}^{d}t_{i,j}^{n_{i,j}(\Sigma)}\frac{1}{|\text{Aut} (\Sigma)|}\prod_{q=1}^{k}\delta_{\ell_{q}(\Sigma),\ell_{q }},
		\end{equation}
		where $n_{ij}(\Sigma)$ is the number of 2-cells with boundaries of lengths $i$ and $j$  used in the gluing of the map $\Sigma$.
		It turns out that these formal series are precisely the genus expansion terms mentioned in the previous section, that is
		
		\begin{equation*}
			\mathcal{T}_{\ell_{1},...,\ell_{k}}=  \sum_{g=0}^{\infty}\left(\frac{N}{t}\right)^{2-2g -k}\mathcal{T}^{g}_{\ell_{1},...,\ell_{k}}
		\end{equation*}
		and 
		\begin{equation*}
			W_{k}(x_{1},..,x_{k}) = \sum_{g=0}^{\infty}\left(\frac{N}{t}\right)^{2-2g -k}W^{g}_{k}(x_{1},...,x_{k}) =\sum_{g=0}^{\infty}\left(\frac{N}{t}\right)^{2-2g -k}\sum_{\ell=0}^{\infty} \frac{\mathcal{T}^{g}_{k}}{x_{1}^{\ell_{1}+1}...x_{k}^{\ell_{k}+1}}.
		\end{equation*}
		For a detailed explanation  and proof of this fact see \cite{blobbed1}.

		\subsection{The Schwinger-Dyson equations}\label{SDE's}
		The Schwinger-Dyson equations (SDE's) provide a powerful method to analyze random matrix models. They were first introduced by Migdal in \cite{Migdal}. In our context they are the consequence of matrix integrals of the total derivative vanishing. The SDE's of the bi-tracial matrix model (\ref{multi trace ensemble})
		can be derived as follows. Consider the following relation,
		
		\begin{equation}\label{SDEr}
			\sum_{i,j} \int_{\mathcal{H}_{N}} \frac{\partial}{\partial H_{ij}} \left(\left(H^{\ell_1} \right)_{ij} \prod_{m=2}^{n}\tr H^{\ell_m} e^{-\tr \tilde{V}(H)}\right)dH = 0,
		\end{equation}
		where $\left(H^{\ell_1} \right)_{ij}$ is the $ij$-th entry of the matrix power $H^{\ell_1}$, and the partial derivative $\frac{\partial}{\partial H_{ij}}$ satisfies
		\begin{equation}\label{pdr}
			\frac{\partial }{\partial H_{ij}}\left( H_{pq}\right)  = \delta_{ip} \delta_{jq}.
		\end{equation}
		One can prove the following properties,
		
		\begin{equation}\label{pd1}
			\frac{\partial }{\partial H_{ij}}\left(H^{\ell_1} \right)_{ij} = \sum_{k=0}^{\ell_{1}-1} \left(H^k\right)_{ii}\left(H^{\ell_1-k-1}\right)_{jj}
		\end{equation}
		and
		
		\begin{equation}\label{pd2}
			\frac{\partial }{\partial H_{ij}}\left(\prod_{m=2}^{n}\tr H^{\ell_m}\right) = \sum_{r=2}^{n}\ell_r \left(H^{\ell_r} \right)_{ji} \prod_{\substack{m=2 \\ m\neq r}}^{n}\tr H^{\ell_m}.
		\end{equation}
		Now, using the Leibniz product rule in (\ref{SDEr}) and the relations (\ref{pd1}) and (\ref{pd2}), we find that
		\begin{align*}
			&\sum_{k=0}^{\ell_{1}-1} \langle \tr H^k \tr H^{\ell_{1}-k-1}  \prod_{m=2}^{n}\tr H^{\ell_m} \rangle + \sum_{r=2}^{n} \ell_r \langle  \tr H^{\ell_{1}+\ell_r-1}  \prod_{\substack{m=2, \\ m\neq r}}^{n}\tr H^{\ell_m} \rangle  \nonumber \\
			&= \Bigg\langle \Bigg(\frac{N}{t} \tr H^{\ell_1+1} +  \sum_{i=3}^d \frac{N}{t} t_i \tr H^{\ell_{1}+i-1} \nonumber \\
			& + \sum_{i,j=1}^d \frac{t_{i,j}}{ij} \bigg( i \tr H^{\ell_{1}+i-1} \tr H^{j} + j \tr H^{i} \tr H^{\ell_{1}+j-1} \bigg) \Bigg) \prod_{m=2}^{n}\tr H^{\ell_m} \Bigg\rangle.
		\end{align*}

		Note the model (\ref{multi trace ensemble}) can be considered as a formal matrix integral or a convergent matrix integral. Either way its moments  will satisfy the nonlinear recursive relationship above. In the case of formal integrals, the equations are valid only order by order in $t$. Applying the genus expansions of moments and cumulants, then collecting terms of the same $N/t$ powers, one finds the following.
		
		\begin{theorem}
			The Schwinger-Dyson equations for the bi-tracial matrix model (\ref{multi trace ensemble}) are :
			\begin{align}\label{SDE}
				&\sum_{k=0}^{\ell_{1}-1} \left( \sum_{h=0}^{g} \sum_{J\subset L} \mathcal{T}_{k,J}^{(h)} \mathcal{T}_{\ell_{1}-k-1,L\setminus J}^{(g-h)} +  \mathcal{T}_{k,\ell_{1}-k-1,L}^{(g-1)} \right)  + \sum_{m=2}^{n} \ell_r   \mathcal{T}_{\ell_{1}+\ell_{r}-1,L\setminus \{ r\}}^{(g)}   \nonumber \\
				&= \mathcal{T}_{\ell_{1}+1,L}^{(g)}  + \sum_{i=3}^{d} t_{i} \mathcal{T}_{\ell_{1}+i-1,L}^{(g)} \nonumber \\
				&+\sum_{i,j=1}^{d}\frac{t_{i,j}}{ij} \left( \sum_{h=0}^{g} \sum_{J\subset L} \left(  i \, \mathcal{T}_{\ell_{1}+i-1,J}^{(h)} \mathcal{T}_{j,L\setminus J}^{(g-h)} + j \, \mathcal{T}_{\ell_{1}+j-1,J}^{(h)} \mathcal{T}_{i,L\setminus J}^{(g-h)} \right) + i \, \mathcal{T}_{\ell_{1}+i-1,j,L}^{(g-1)} + j \, \mathcal{T}_{\ell_{1}+j-1,i,L}^{(g-1)} \right),
			\end{align}
			where $L=\{ \ell_{2},\ell_{3},\cdots,\ell_{n} \}$.
		\end{theorem}
		The coefficient of $N/t$, when the number of boundaries $n$ is equal to 1, is the main object of study in the following section.

		\subsection{The spectral curve}
		When $g = 0$ and $n=1$, equation (\ref{SDE}) becomes
		\begin{equation*}
			\sum_{k=0}^{\ell -1} \mathcal{T}_{k}^{0} \mathcal{T}_{\ell-k-1}^{0} =  \mathcal{T}_{\ell + 1}^{0} + \sum_{i=3}^{d}t_i\mathcal{T}_{\ell+i-1}^{0} + \sum_{i,j=1}^{d} \frac{t_{i,j}}{ij}\left(i \, \mathcal{T}^{0}_{\ell+i-1} \mathcal{T}^{0}_{j}+j \, \mathcal{T}^{0}_{\ell+j-1}\mathcal{T}^{0}_{i} \right).
		\end{equation*}
		This is significantly more complicated than the single trace case. To simplify the notation we write it as 
		\begin{equation}\label{first loop}
			\sum_{k=0}^{\ell -1} \mathcal{T}_{k}^{0} \mathcal{T}_{\ell-k-1}^{0} = \tilde{t}_{2}\mathcal{T}_{\ell + 1}^{0} +\tilde{t}_{1} \mathcal{T}^{0}_{\ell} + \sum_{i=3}^{d}\tilde{t}_{i}\mathcal{T}_{\ell+i-1}^{0},
		\end{equation}
		where  the bi-tracial terms become a combination of moments and coupling constants $t_{j}$, all inside the new $\tilde{t}_{j}$'s as follows:
		\begin{equation}\label{new couplings}
			\tilde{t}_{i} = \delta_{2,i}+\sum_{j=3}^{d} \delta_{i,j} t_i + 2 \sum_{j=1}^{d} \frac{t_{i,j}}{j}  \mathcal{T}^{0}_{j}.
		\end{equation}
		Note that these new coupling constants are formal series in the old coupling constants since they are composed of the old coupling constants and $\mathcal{T}^{0}_{\ell}$ for various $\ell$, which are also formal series in the original coupling constants.
		Next we multiply equation (\ref{first loop}) by $1/x^{\ell+1}$ and sum from $\ell = 0$ to infinity. The resulting equation is 
		
		\begin{equation*}
			\sum_{\ell=0}^{\infty} \sum_{k=0}^{\ell -1} \frac{\mathcal{T}_{k}^{0} \mathcal{T}_{\ell-k-1}^{0}}{x^{\ell+1}} = \sum_{\ell=0}^{\infty} \tilde{t}_{2} \frac{\mathcal{T}_{\ell + 1}^{0}}{x^{\ell+1}}  +\sum_{\ell=0}^{\infty} \tilde{t}_{1} \frac{\mathcal{T}^{0}_{\ell}}{x^{\ell+1}}  + \sum_{\ell=0}^{\infty} \sum_{i=3}^{d}\tilde{t}_{i} \frac{\mathcal{T}_{\ell+i-1}^{0}}{x^{\ell+1}}.
		\end{equation*}
		Then the equation can be written as  
		\begin{equation}\label{spectral curve}
			W_{1}^{0}(x)^{2} = S'(x)W_{1}^{0}(x) - P_{1}^{0}(x),
		\end{equation}  
		where 
		\begin{equation*}
			W_{1}^{0}(x) = \sum_{\ell=0}^{\infty}\frac{\mathcal{T}_{\ell}^{0}}{x^{\ell+1}},
		\end{equation*}
		\begin{equation}\label{Polynomial}
			P_{1}^{0}(x) = t + \sum_{j=2}^{d}\sum_{\ell = 0}^{j-2}\tilde{t}_{j} \mathcal{T}^{0}_{j-\ell-2}x^{\ell},
		\end{equation}
		and
		\begin{equation}\label{potential}
			S(x) = \frac{1}{2} \tilde{t}_{2} x^{2}  +\tilde{t}_{1} x+ \sum_{j=3}^{d}\frac{\tilde{t}_{j}}{j} x^{j}.
		\end{equation}
		
		By solving  the quadratic equation (\ref{spectral curve}), one can find the resolvent function  
		\begin{equation}\label{resolvent}
			W_{1}^{0}(x) = \frac{1}{2}\left(S'(x) - \sqrt{S'(x)^{2} - 4 P_{1}^{0}(x)}\right).
		\end{equation}
		
		Equation (\ref{spectral curve}) is commonly referred to as the \textit{spectral curve}. The solution $W_{1}^{0}(x)$, called the resolvent, can be used to give us the limiting eigenvalue density function of the random matrix associated to the given model. It is well-known  that this relationship is given by the Stieltjes transform
		\begin{align*}
			W^{0}_{1}(x) &=  \sum_{\ell=0}^{\infty}\frac{\mathcal{T}_{\ell}^{0}}{x^{\ell+1}} =  \sum_{\ell=0}^{\infty}\frac{\lim_{N \rightarrow \infty}\langle \frac{1}{N} \tr H^{\ell}\rangle }{x^{\ell+1}}\\ &= \lim_{N \rightarrow \infty}\langle \frac{1}{N} \tr \sum_{\ell=0}^{\infty}\frac{H^{\ell}}{x^{\ell+1}} \rangle = \lim_{N \rightarrow \infty} \frac{1}{N}\langle\tr \frac{1}{x-H} \rangle \\
			&= \text{p.v.}\int_{\text{supp}\rho}\frac{\rho(y)}{x-y}dy,
		\end{align*}
		where we define $\rho(y)$ as the limiting eigenvalue density function of the model. Thus, applying the inverse Stieltjes transform tells us that
		\begin{equation*}
			\rho(x) = -\frac{1}{\pi t} \Im W_{1}^{0}(x).
		\end{equation*} 
		Consider for example the case that all coupling cosntants are zero except for $t=1$. The resulting model is the Gaussian Unitary Ensemble and 
		\begin{equation*}
			W_{1}^{0}(x) = \frac{1}{2}(x-\sqrt{x^2 -4}).
		\end{equation*} 
		One can also derive that $\text{supp}\rho = [-2,2]$. So
		\begin{equation*}
			\rho(x) = \frac{1}{2\pi}\sqrt{4-x^{2}}_{[-2,2]},
		\end{equation*}
		which is Wigner's semicircle distribution.
		
		\section{The 1-cut lemma}
		The following theorems for unstable stuffed maps are based on the results of \cite{Second paper} and is a generalization of a famous result by the same name discussed in \cite{Eynard2018}. Ideally one would like to know how the discriminant in equation (\ref{resolvent}) factors. This will determine the number of connected components in the support of the limiting spectral distribution and help with finding a nice closed form.
		
		\begin{lemma}\label{vertices bound}
			Give an unstable planar map  with at least one vertex glued from 
			\begin{itemize}
				\item one boundary of length $\ell\geq 1$ i.e. rooted 2-cells with the topology of the disc
				\item $n_{j}$ $j$-gons for $3 \leq j \leq d$
				\item some finite set of 2-cells with the topology of a cylinder,
			\end{itemize} then the number of vertices is greater than or equal to two.
		\end{lemma}

		\begin{proof}
			Suppose we want to construct a planar map with one vertex and one of the 2-cells used in the gluing has the topology of the cylinder. Then since there is only one vertex both faces of the cylinder must be glued together in some manner to meet this requirement. Regardless of the configuration of the gluing, one will have created a handle by gluing the faces of the cylinder together. This contradicts the assumption that the map is planar. Hence, no planar map with one vertex can be glued from a 2-cell with the topology of the cylinder. Thus, an unstable planar map with one vertex must be glued from polygons only. However, we know from equation 1.2.1 of \cite{Eynard2018} that the number of vertices of such a map is equal to
			\begin{equation}
				1+\frac{\ell}{2}+\frac{1}{2}\sum_{j=3}^{d}(j-2)n_{j} \geq 2.
			\end{equation} 
			This completes the proof.
		\end{proof}
		
		\begin{theorem}[1-Cut/Brown's Lemma]
			There exists formal power series  $\alpha$ and $\gamma^{2}$, as well as a polynomial  $M(x)$, such that 
			\begin{equation*}
				\alpha = O(t), \qquad \gamma^{2} = t + O(t^{2}), \qquad M(x)= \frac{S'(x)-\tilde{t}_{1}}{x} + O(t),
			\end{equation*}
			and 
			\begin{equation}\label{discrim}
				\sqrt{S'(x)^{2} - 4 P_{1}^{0}(x)} = M(x)\sqrt{(x-a)(x-b)},
			\end{equation}
			with $a = \alpha +2\gamma$ and  $b = \alpha - 2\gamma$.
		\end{theorem}
		Brown's lemma tells us that the models we are interested in will always have a single cut solution. In general it is possible to find the number of connected components of  support of $\rho$. This will depend on values of the coupling constants as well as the degree and structure of $S(x)$. Note that the statement of this theorem differs slightly from the single trace case.
		
		\begin{proof}
			This proof follows closely that of Lemma 3.1.1 of \cite{Eynard2018}. Our initial goal is to first show that the zeros of our potential are up to some order in $t$  close to the zeros of the discriminant $S'(x)^{2} - 4 P_{1}^{0}(x)$. What order in $t$ is key to the existence of the single cut solution.
			
			Recall definition (\ref{moments}) of $\mathcal{T}^{0}_{\ell}$. For $\ell = 0$, $\mathcal{T}^{0}_{\ell} = t$. However, when $\ell \geq 1$,  Lemma \ref{vertices bound} is applicable, and says that  $\mathcal{T}^{0}_{\ell}$ is a formal series starting at a $t^{2}$ term. In particular, using definition (\ref{new couplings}) it is clear that $\tilde{t}_{1}=\mathcal{O}(t^{2})$.
			
			Now recall the equation (\ref{Polynomial}). Using the above information about  $\mathcal{T}^{0}_{\ell}$, we may write it as 
			\begin{align*}
				P^{0}_{1}(x) &=t + \sum_{j=2}^{d}\sum_{k = 0}^{j-2}\tilde{t}_{j} \mathcal{T}^{0}_{j-k-2}\,x^{k}\\
				&= t\left(1 + \sum_{j=2}^{d} \tilde{t}_{j} x^{j-2}\right) + \mathcal{O}(t^{2}) \\
				&= \frac{S'(x)-\tilde{t}_{1}}{x}t + \mathcal{O}(t^{2}),
			\end{align*}
			where the last line follows from rearranging equation (\ref{potential}).
			
			Let $x_{1},x_{2},...,x_{d-1}$ be the roots of $S'(x)-\tilde{t}_{1}$. For simplicity let $x_{i}$ be a root such that $S''(x_{i}) \not = 0$. Let $x$ be in the $t$-neighborhood of $x_{i}$. Then we may write 
			
			\begin{align*}
				S'(x)-\tilde{t}_{1} &= S'(x_{i}) -\tilde{t}_{1}+ (x-x_{i})S''(x_{i}) + \frac{1}{2}(x-x_{i})^{2}S'''(x_{i}) +...\\
				&= (x-x_{i})S''(x_{i}) + \mathcal{O}(t^{2}).
			\end{align*}
			This allows us to write 
			\begin{align*}
				S'(x)^{2}-4P_{1}^{0}(x)&= (x-x_{i})^{2}(S''(x_{i}))^{2} - 4 P_{1}^{0}(x) + \mathcal{O}(t^{2})\\
				&= S''(x_{i})^{2}\left((x-x_{i})^{2} - \frac{4 P_{1}^{0}(x)}{(S''(x_{i}))^{2}}\right)+\mathcal{O}(t^{2})\\
				&= S''(x_{i})^{2}\left(x-x_{i}- \frac{2 \sqrt{P_{1}^{0}(x)}}{S''(x_{i})}\right)\left(x-x_{i}+ \frac{2 \sqrt{P_{1}^{0}(x)}}{S''(x_{i})}\right)+\mathcal{O}(t^{2}).
			\end{align*}
			
			This shows that the zeros come in pairs $[a_{i},b_{i}]$ centered around $x_{i}$ up to some order in $t$. The key question now is to what order. We have two cases to consider. Without loss of generality let $x_{1}=0$ (zero will always be a root since $S'(x)-\tilde{t}_{1}$ lowest order term is of degree two), then 
			\begin{equation*}
				P_{1}^{0}(0) = tS''(0)+ \mathcal{O}({t^{2}}),
			\end{equation*}
			so 
			\begin{equation*}
				P_{1}^{0}(x)- P_{1}^{0}(0) = \mathcal{O}(t).
			\end{equation*}
			If on the other hand 
			$x_{i}\not =0$, then 
			\begin{equation*}
				P_{1}^{0}(x_{i}) = 0+ \mathcal{O}({t^{2}}),
			\end{equation*}
			so 
			\begin{equation*}
				P_{1}^{0}(x)- P_{1}^{0}(x_{i}) = \mathcal{O}(t^{2}).
			\end{equation*}
			In the first case this implies that the roots of $	S'(x)^{2}-4P_{1}^{0}(x)$ can be written as 
			\begin{equation}\label{roots one}
				a:=a_{1}= 2 \sqrt{t} + o(\sqrt{t}), \qquad b :=b_{1} =- 2 \sqrt{t}+ o(\sqrt{t})
			\end{equation}
			and for the second case
			\begin{equation}\label{roots two}
				a_{i} = x_{i}+ \mathcal{O}(t), \qquad b_{i} = x_{i}+ \mathcal{O}(t).
			\end{equation}
			
			As candidates for variables of the same name in the statement of the theorem define $\alpha = \frac{1}{2}(a+b)$ and $\gamma = \frac{1}{4}(a-b)$, then
			\begin{equation*}
				\gamma^{2} = t + \mathcal{O}(t^{2}), \qquad \alpha = \mathcal{O}(t).
			\end{equation*}
			
			We now just need to show the last equality in the statement of the theorem for our choices of $a$ and $b$. That is, we want to show that the discriminant $S'(x)^{2}-4P_{1}^{0}(x)$ only has one pair of simple zeros. This implies that we can pullout all the double zeros, giving us the polynomial $M(x)$.
			
			We know from Corollary 3.1.1 of \cite{Second paper}, that the number of unstable planar maps with $v$ vertices and a boundary of length  $\ell$ is finite. Thus, $W^{0}_{1}(x)$ is a formal power series in $t$ and may be written as 
			\begin{equation*}
				W_{1}^{0}(x) = \frac{t}{x} + \sum_{v=2}^{\infty}t^{v}c_{v}(x),
			\end{equation*} 
			where $c_{v}(x)$ is some Laurent polynomial of at least order two in $x$ with coefficients formed in terms of the coupling constants. Thus, for any positively oriented contour $\mathcal{C}$, integrating term-wise we know that
			\begin{equation}\label{contour of resolvent}
				\frac{1}{2 \pi i}\oint_{\mathcal{C}} W^{0}_{1}(z) dz= \begin{cases} 
					t & \text{if zero is in}\,\, \mathcal{C} \\
					0& \text{otherwise}
				\end{cases}.
			\end{equation}  
			
			Consider now a contour centered around one of the $x_{i} \not = 0$, which for suitably small $t$ contains $a_{i}$ and $b_{i}$ by equation (\ref{roots two}). Let us compute the resolvent using (\ref{resolvent}) and Cauchy's integral theorem 
			\begin{align*}
				\frac{1}{2 \pi i}\oint_{\mathcal{C}} W^{0}_{1}(z) dz&=	\frac{1}{2 \pi i}\oint_{\mathcal{C}}\frac{1}{2}\left(S'(z) - \sqrt{S'(z)^{2} - 4 P_{1}^{0}(z)}\right)dz\\
				&= -\frac{1}{4 \pi i}\oint_{\mathcal{C}} \sqrt{S'(z)^{2} - 4 P_{1}^{0}(z)}dz\\
				&=-\frac{1}{4 \pi i}\oint_{\mathcal{C}}\sqrt{(z-a)(z+b)(z-a_{2})(z+b_{2})\cdots(z+b_{d-1})}dz.\\
			\end{align*}
			By the Taylor series expansion of the function 
			\begin{equation*}
				\sqrt{x-a_{i}} = 	\sqrt{x-x_{i}} +\frac{(x_{i}-a_{i})}{2\sqrt{x-x_{i}}} -\frac{(x_{i}-a_{i})^{2}}{8(x-x_{i})^{3/2}} + o((x_{i}-a_{i})^{2}),
			\end{equation*}
			we have that 
			\begin{equation*}
				\sqrt{(x-a_{i})(x-b_{i})} = (x - x_{i}) + \frac{1}{2}(2x_{i}-a_{i}-b_{i}) -\frac{(a_{i}-b_{i})^{2}}{8(x-x_{i})} + o((x_{i}-a_{i})^{2}).
			\end{equation*}
			
			Suppose that $a_{i}-b_{i}\not = 0$, then by the above expansion the integrand will have a pole at $x_{i}$ whose term is $\mathcal{O}(t^{2})$. This contradicts equation (\ref{contour of resolvent}).
			
			Thus, we have shown that $a_{i} =b_{i}$ for $i>1$. In other words, for sufficiently small $t$ the polynomial $S'(x)^{2}-4P_{1}^{0}(x)$ has only one pair $a$ and $b$ of simple zeros.  This shows that the polynomials factors as $M(x)^{2}(x-a)(x-b)$ for some range of the coupling constants. This proves the last statement of the theorem and completes the proof.

		\end{proof}

		\section{The resolvent}
		In this section we will focus on finding the resolvent $W_{1}^{0}(x)$ for several formal bi-tracial matrix models that are type $(1,0)$ Dirac ensembles.
		\subsection{The Zhukovsky transform}
		The form of the discriminant in equation (\ref{discrim}) motivates the use of the Zhukovsky transform
		\begin{equation*}
			x(z)
			= \frac{a+b}{2} + \frac{a-b}{4}\left(z + \frac{1}{z}\right)
		\end{equation*}
		with an inverse
		\begin{equation*}
			z = \frac{1}{2\gamma}\left(x -\alpha + \sqrt{(x-\alpha)^{2}-4\gamma^{2}}\right).
		\end{equation*}
		The Zhukovsky transform maps the x-plane minus a line segment to the exterior of the unit disk in the z-plane. 
		It also has the following useful identity
		\begin{equation*}
			\sqrt{(x(z)-a)(x(z)-b)} = \frac{a-b}{4}\left(z-\frac{1}{z}\right).
		\end{equation*}
		
		We now borrow Theorem 3.1.1 from \cite{Eynard2018}, which applies to our models as well.
		\begin{theorem}\cite{Eynard2018} \label{u's}
			For the formal power series $\alpha$ and $\gamma^{2}$ as mentioned above, we have the expansions 
			\begin{equation*}
				S'(x(z)) = \sum_{k=0}^{d-1}u_{k}(z^{k}+z^{-k})
			\end{equation*}
			and 
			\begin{equation*}
				W_{1}^{0}(x(z)) = \sum_{k=0}^{d-1}u_{k}z^{-k},
			\end{equation*}
			with $u_{0} = 0$ and $u_{1} = t/\gamma$.
		\end{theorem}

		Notice that the theorem implies that since we have  $S(x)$,  in theory we should be able to compute the coefficients of $W_{1}^{0}(x(z))$ and then transform back to get $W_{1}^{0}(x)$. Computing the coefficients $u_{k}$ is much more involved in our models than in the single trace cases seen in \cite{Eynard2018}. We will show how to find them in various examples.  
		\subsection{Moments}
		In general one may apply Lagrange's inversion formula to compute the moments and the support of $\rho (x)$ for a given model.  To see this one first observes that moments can be extracted from the resolvent generating function via the following contour integral
		\begin{equation*}
			\mathcal{T}^{0}_{\ell} = -\frac{1}{2\pi i}\oint_{\mathcal{C}} x^{\ell}W_{1}^{0}(x)dx.
		\end{equation*}
		We then apply the Zhukovsky transform to get
		\begin{equation*}
			\mathcal{T}^{0}_{\ell} = -\frac{1}{2\pi i}\oint_{\mathcal{C}} x(z)^{\ell}W_{1}^{0}(x(z))x'(z)dz.
		\end{equation*}
		Expanding one can find the general formula.
		\begin{corollary} \cite{Eynard2018} \label{moments}
			The $\ell$-th moment of $\rho(x)dx$ is 
			\begin{equation*}
				\mathcal{T}^{0}_{\ell} = \sum_{i+j < \ell, i < j < i+d}\frac{(j-i)\ell!}{(i+1)!(j+1)! (\ell-1-i-j)!} \alpha^{\ell - 1-i-j} \gamma^{i+j+2} u_{j-i},
			\end{equation*}
			where the $u_{j-i}$'s are the coefficients from theorem \ref{u's} that are determined by the potential. 
		\end{corollary}
		This formula looks the same as for single trace models, but it is important to note that the $u_{k}$'s contain other moments of the model because of the bi-tracial terms. Because of this, this formula now gives us a system of nonlinear equations to solve for moments. 
		
		As a side note, if one wishes to compute $\gamma$ or $\alpha$, the Lagrange inversion formula can be applied to recover them in terms of the coupling constants. Just like in the single trace case even though we treated moments as  just  formal series, they are in fact algebraic functions of $\alpha$ and $\gamma$, which are algebraic functions of the coupling constants. Thus, they have a finite number of singularities and are convergent in some sufficiently small disc. These singularities are a beautiful artifact of these models that is fundamental for the connection to Liouville quantum gravity, as we will see. The characterization of these singularities is not quite understood for multi-trace models in general, and may be an interesting phenomenon to study.
		
		In the following sections we will set	$t=1$ but the calculations that follow can be carried out with  $t \not =1$.	
		
		\subsection{The quartic model}\label{quartic analysis}
		Consider the  potential 
		\begin{equation*}
			V(D) = \frac{t_{2}}{4} D^{2} + \frac{t_{4}}{8} D^{4}.
		\end{equation*}
		Then the trace of the potential can be written in terms of the Hermitian matrix $H$:
		\begin{equation*}
			\tr \tilde{V}(H) =\frac{N}{2}  t_{2} \tr H^{2} + \frac{N}{4}t_{4}\tr H^{4} +\frac{3}{4} t_{4} \tr H^{2} \tr H^{2}.
		\end{equation*}
		Note that we neglect terms with traces of odd powers due to their lack of contribution to the large $N$ limit. The derivative of the potential in the spectral curve (\ref{spectral curve}) is given by
		\begin{equation*}
			S'(x) = (t_{2} + 3\mathcal{T}_{2}^{0} t_{4}) x +t_{4}x^{3}.
		\end{equation*}
		For this model $\mathcal{T}^{0}_{\ell}$ counts the number of planar gluings of quandrangles and 2-cells with two boundaries of lengths two, with one polygon boundary of length $\ell$. If $\ell$ is odd no such gluings exist so the generating function is zero. Hence, we focus on even values of $\ell$. 
		
		We next transform the resolvent and find that 
		
		\begin{equation*}
			W_{1}^{0}(x(z)) = \frac{1}{\gamma z}+t_{4}\gamma^{3}\frac{1}{z^{3}}.
		\end{equation*}
		Transforming back, we arrive at 
		\begin{equation*}
			W_{1}^{0}(x) =	\frac{1}{2}\left(  (t_{2}+3\mathcal{T}_{2}^{0} t_4)x +t_{4}x^{3} - t_{4}(x^{2}-\gamma^{2}+\frac{t}{t_{4}\gamma^{2}} )\sqrt{x^2 -4\gamma^2} \right).
		\end{equation*}
		Thus, the limiting eigenvalue density function is 
		\begin{equation*}
			\rho(x) = -\frac{1}{\pi t} \Im W_{1}^{0}(x) = \frac{1}{2\pi}\left( t_{4}(x^{2}-\gamma^{2})+\frac{1}{\gamma^{2}} \right)\sqrt{4\gamma^2-x^2}_{[-2\gamma,2\gamma]}.
		\end{equation*}
		
		One would like to be able to find the value of $\gamma$ in terms of the coupling constants. To do this, we start by finding the transformed coefficients of $S'$ : 
		\begin{equation*}
			S'(x(z)) =(t_{2} + 3\mathcal{T}_{2}^{0} t_{4})(\alpha +\gamma(z+1/z)) +t_{4}(\alpha +\gamma(z+1/z))^{3}. 
		\end{equation*}
		Expanding, we find that we may write the coefficients of $S'(x(z))$ in Theorem \ref{u's} as follows:
		\begin{align*}
			u_{0} &= \alpha(t_{2}+3\mathcal{T}_{2}^{0} t_4) + t_{4}(\alpha^{3}+6\alpha \gamma^{2})\\
			u_{1}&= (t_{2}+3\mathcal{T}_{2}^{0} t_4)\gamma +t_{4}(3 \alpha^{2}\gamma +3\gamma^{3} )\\
			u_{2} &= 3t_{4}\alpha \gamma^{2}\\
			u_{3}&= t_{4}\gamma^{3}.
		\end{align*}
		Using the same theorem, we deduce that 
		
		\begin{align*}
			0&= u_{0} = \alpha((t_{2}+3\mathcal{T}_{2}^{0} t_4) + t_{4}(\alpha^{2}+6 \gamma^{2}))
		\end{align*}
		and 
		\begin{align*}
			\frac{1}{\gamma} &= u_{1} = \gamma((t_{2}+3\mathcal{T}_{2}^{0}t_4) +t_{4}(3 \alpha^{2} +3\gamma^{2} )).
		\end{align*} 
		By the one-cut lemma as $t$ goes to zero, so does $\alpha$ and $\mathcal{T}_{2}^{0}$ by definition. Thus the factor $((t_{2}+3\mathcal{T}_{2}^{0}t_4)) + t_{4}(\alpha^{2}+6 \gamma^{2}))$ is nonzero when $t_{2} \neq 0$ order by order in $t$. Hence $\alpha =0$ in order for this equation to hold on each term of the formal sum.
		This reduces the second equation to
		\begin{equation*}
			3t_{4}\gamma^{4}+ (t_{2}+3\mathcal{T}_{2}^{0} t_4)\gamma^{2}-1 =0
		\end{equation*}
		Using the formula for the second moment  from Corollary \ref{moments}, we have
		\begin{equation*}
			3\,{t}_4^2{\gamma}^{8}+6t_{4} \gamma^4 + t_{2} \gamma^{2} -1=0.
		\end{equation*}
		Thus we are left with an equation that relates $\gamma$ to the coupling constants, which can be solved symbolically using software such as Maple or Mathematica.
		
		Note that if we travel along the curve $t_{2} = 1- 3 \mathcal{T}_{2}^{0}t_{4}$ the derivative of the potential curve becomes 
		\begin{equation*}
			S'(x) = x-t_{4}x^3,	
		\end{equation*}
		which is the same as the quartic Hermitian matrix model \begin{equation*}
			\int_{\mathcal{H}_{N}} e^{-\frac{N}{2}\tr H^{2} - \frac{N}{4} t_{4}\tr H^{4}} dH.
		\end{equation*}
		See chapter 3 of \cite{Eynard2018} for details about this model. Thus, both models at these specifications have the same resolvent $W^{0}_{1}(x)$. This shall be important in later sections.  We shall make similar conclusions about the rest of the models examined in the following two subsections.
		
		\subsection{The hexic model}
		Consider the potential 
		\begin{equation*}
			V(D) =\frac{t_2}{4}  D^2 + \frac{t_4}{8}  D^4 + \frac{t_6}{12} D^6.
		\end{equation*}
		For the spectral curve  (\ref{spectral curve}) we get 
		\begin{equation*}
			S(x) = {\frac {{ t_2}\,{x}^{2}}{2}}+{t_4}\, \left( {\frac {{x}^{4}}{4}}+
			{\frac {3\,{ \mathcal{T}^{0}_{2}}\,{x}^{2}}{2}} \right) +{t_6}\, \left( {\frac {{
						x}^{6}}{6}}+{\frac {5\,{\mathcal{T}^{0}_{2} }\,{x}
					^{4}}{2}}+{\frac {5\,{ \mathcal{T}^{0}_{4}}\,{x}^{2}}{2}} \right),
		\end{equation*}
		and thus
		\begin{equation*}
			S'(x) ={t_6}\,{x}^{5}+ \left( 10\,{ \mathcal{T}^{0}_{2}}\,{t_6}+{ t_4} \right) {x}^
			{3}+ \left( 5\,{ \mathcal{T}^{0}_{4}}\,{ t_6}+3\,{\mathcal{T}^{0}_{2}}\,{ t_4}+{ t_2}
			\right) x.	
		\end{equation*}
		For this model, if $\ell$ is odd the moment is zero by the same $H \rightarrow -H$ symmetry as in the quartic case. We then apply the Zhukovsky transform to obtain
		\begin{equation*}
			\alpha^5 t_6 + (t_4 + t_6 (20 \gamma^2 + 10 m_2)) \alpha^3 + (t_2 + t_4(6\gamma^2 + 3 m_2) + t_6 (30 \gamma^4 + 60\gamma^2 m_2 + 5 m_4 ))\alpha = u_{0} = 0.
		\end{equation*}
		By the same argument as in the quartic case, we can deduce $\alpha = 0$. Thus, we find
		
		\begin{align*}
			\frac{1}{\gamma}&= u_{1}= t_2 \gamma + t_4 (3\gamma^3 + 3\gamma m_2) + t_6(10\gamma^5 + 30\gamma^3 m_2 + 5\gamma m_4)\\
			u_{2} &= 0\\
			u_{3}&= t_4 \gamma^3 + t_6 (5\gamma^5 + 10\gamma^3 m_2)\\
			u_{4}&= 0\\
			u_{5}&= t_6 \gamma^5.
		\end{align*}
		Using the Corollary \ref{moments}, we find that 
		\begin{align*}
			m_{2} &= -{\frac {{{\gamma}}^{2} \left( 5\,{t_6}\,{{\gamma}}^{6}+{t_4}
					\,{{\gamma}}^{4}+1 \right) }{10\,{ t_6}\,{{\gamma}}^{6}-1}}\\
			m_{4}&= 2\,{{\gamma}}^{4}+3\,{{\gamma}}^{5} \left( {t_4}\,{{\gamma}}^{3}
			+{t_6}\, \left( 5\,{{\gamma}}^{5}-10\,{\frac {{{\gamma}}^{5}
					\left( 5\,{t_6}\,{{\gamma}}^{6}+{t_4}\,{{\gamma}}^{4}+1
					\right) }{10\,{t_6}\,{{\gamma}}^{6}-1}} \right)  \right) +{t_6
			}\,{{\gamma}}^{10}.
		\end{align*}
		This gives the limiting eigenvalue density function as
		\begin{align}\label{rho hexic}
			\rho(x) &= -\frac{1}{\pi t} \Im W_{1}^{0}(x) =\frac{1}{2 \pi} \left({t_6}\,{x}^{4}+{\frac {{x}^{2}}{{\gamma}} \left( {\frac {{t_4}
						\,{{\gamma}}^{3}+{t_6}\, \left( 5\,{{\gamma}}^{5}+10\,{{\gamma}}
						^{3}{ m_2} \right) }{{{\gamma}}^{2}}}-3\,{ t_6}\,{{\gamma}}^{3}
				\right) }\right. \nonumber\\ 
			&\left.+{\frac {{{\gamma}}^{-1}-{t_4}\,{{\gamma}}^{3}-{t_6}
					\, \left( 5\,{{\gamma}}^{5}+10\,{{\gamma}}^{3}{m_2} \right) +{
						t_6}\,{{\gamma}}^{5}}{{\gamma}}}
			\right)\sqrt{4\gamma^2-x^2}_{[-2\gamma, 2\gamma]},
		\end{align}
		where $\gamma$ can be found as the solution to the $1/\gamma = u_{1}$ relation for given $t_{2},t_{4}$, and $t_{6}$.
		
		Similarly as in the quartic ensemble, when one travels along the hypersurface $5\,{ \mathcal{T}^{0}_{4}}\,{ t_6}+3\,{\mathcal{T}^{0}_{2}}\,{ t_4}+{ t_2} = 1$, we have that  the derivative of the potential becomes
		\begin{equation*}
			S'(x)={t_6}\,{x}^{5}+ \left( 10\,{ \mathcal{T}^{0}_{2}}\,{t_6}+{ t_4} \right) {x}^
			{3}+  x.
		\end{equation*}
		Next, one may set $10\,{ \mathcal{T}^{0}_{2}}\,{t_6}+{ t_4}$ equal to some desired value to reduce the derivative of the potential, and therefore $W_{1}^{0}(x)$, to being the same as the hexic Hermitian matrix model
		\begin{equation*}
			\int_{\mathcal{H}_{N}} e^{-\frac{N}{2}\tr H^{2} - \frac{N}{4} t_{4}\tr H^{4} -\frac{N}{6} t_{6} \tr H^{6}} dH.
		\end{equation*}

		\subsection{The cubic model}
		Consider the potential 
		\begin{equation*}
			V(D) =\frac{t_2}{4} D^2 + \frac{t_3}{6} D^3,
		\end{equation*}
		which gives us
		\begin{equation*}
			S(x) = {\frac {{t_2}\, \left( 2\,{ m_1}\,x+{x}^{2} \right) }{2}}+{\frac 
				{{t_3}\, \left( 3\,{m_1}\,{x}^{2}+{x}^{3}+3\,x{m_2} \right) 
				}{3}}
		\end{equation*}
		in the spectral curve equation (\ref{spectral curve}). Thus, 
		\begin{equation*}
			S'(x) ={t_3}\,{x}^{2}+ \left( 2\,{t_3}\,{m_1}+{t_2} \right) x+{
				t_2}\,{m_1}+{t_3}\,{m_2}
			.	
		\end{equation*}
		We then apply the Zhukovsky transform and find that 
		
		\begin{align*}
			0& = u_0 =  {\frac {{t_2}\, \left( 2\,{m_1}+2\,a \right) }{2}}+{\frac {{
						t_3}\, \left( 3\,{\alpha}^{2}+6\,\alpha {m_1}+6\,{{\gamma}}^{2}+3\,{m_2}
					\right) }{3}}\\
			\frac{1}{\gamma}&= {t_2}\,{\gamma}+{\frac {{ t_3}\, \left( 6\,{\gamma}\,\alpha +6\,{m_1}\,{\gamma} \right) }{3}}
			\\
			u_{2} &= t_3 \gamma^2 z^2.
		\end{align*}
		Using Corollary \ref{moments}, we may compute the moments of interest  
		\begin{align*}
			m_{1} &= \alpha + \gamma^{4} t_{3}\\
			m_{2}&= \alpha^{2}+ \gamma^{2}+2\alpha \gamma^{4} t_{3}.
		\end{align*}
		This gives the limiting eigenvalue density function 
		\begin{align*}
			\rho(x) &= -\frac{1}{\pi t} \Im W_{1}^{0}(x) =\frac{1}{ \pi} \left(\frac{1}{\gamma^{2}}+ t_{3}(x - \alpha) \right)\sqrt{(x- \alpha - 2\gamma)(\alpha - 2\gamma -x)}_{[\alpha -2\gamma,\,\alpha +2 \gamma]},
		\end{align*}
		where $\gamma$ and $\alpha$ can be found as the solution to the above relations for given $t_{2}$ and $t_{3}$.
		
		When one travels along $ 2\,{t_3}\,{m_1}+{t_2}  = 1$ and $
		t_2\,{m_1}+{t_3}\,{m_2} = 0$, the derivative of the potential is the same as the cubic Hermitian matrix model
		\begin{equation*}
			\int_{\mathcal{H}_{N}} e^{-\frac{N}{2}\tr H^{2} - \frac{N}{3} t_{3}\tr H^{3}} dH.
		\end{equation*}
		Thus, in these circumstances both models have the same $W_{1}^{0}(x)$.

		\section{(Blobbed) topological recursion}
		Roughly speaking, and   in our context of matrix models,  topological recursion works as follows. Using the resolvent technique one first defines a complex curve (Riemann surface) $\Sigma$, called the spectral curve of the model. One then   constructs a sequence of symmetric mermorphic differential forms $\omega_{g, n} (z_1, \dots, z_n)dz_{1}...d{z_{n}} $ of degree $n$ for $ g \geq 0, n \geq 1, $ on $n$-fold Cartesian products of $\Sigma$. 
		Topological recursion works by induction on the Euler characteristic $ 2-2g -n$ of a surface of genus $g$ with $n$ boundaries. It gives an inductive  formula for all $\omega_{g,n} $, starting with the first two forms $\omega_{0,1}$ and $\omega_{0, 2}$:
		\begin{align*}
			\omega_{g, n+1}(I, z) &=\sum_{\beta_{i}} \operatorname{Res}_{q \rightarrow \beta_{i}} K_{i}(z, q)\Big(\omega_{g-1, n+2}\left(I, q, \sigma_{i}(q)\right)\\
			&+\sum_{\substack{g_{1}+g_{2}=g \\
					I_{1} \uplus I_{2}=I \\
					\left(g_{1}, I_{1}\right) \neq(0, \emptyset) \neq\left(g_{2}, I_{2}\right)}} \omega_{g_{1},\left|I_{1}\right|+1}\left(I_{1}, q\right) \omega_{g_{2},\left|I_{2}\right|+1}\left(I_{2}, \sigma_{i}(q)\right)\Big),
		\end{align*}
		where $I = \{z_{1},...,z_{n}\}$ and $\beta_{i}$ are the ramification points of the ramified covering $x$, defined via $dx(\beta_{i}) = 0$. The recursive kernel $K_{i}(z,q)$ is constructed from the initial data.  In the special case when $\Sigma$ is the projective line, $\omega_{0,2}$ is the {\it Bergmann kernel}
		$$ \omega_{0,2} (z_1, z_2)  =\frac{dz_1 dz_2}{(z_1-z_2)^2}.$$

		As previously mentioned, it was first seen in \cite{blobbed1} that multi-trace matrix models correspond to the generating function of stuffed maps. Such generating functions obeys a generalization of topological recursion \cite{blobbed1}, known as blobbed topological recursion. Recall the generating functions  $W_{k}^{g}(x_{1},..,x_{k})$ of unstable stuffed maps defined in Section \ref{Stuffed maps}. As discussed these generating functions are precisely the resolvent functions' genus expansion terms for multi-trace matrix models. In our case we are dealing with bi-tracial models, whose genus expansion  is proven rigorously in \cite{Second paper} which corresponds to unstable stuffed maps i.e. stuffed maps involving 2-cells with the topology of discs and cylinders. Given the initial generating functions $W_{1}^{0}(x)$ and $W_{2}^{0}(x_{1},x_{2})$, which count connected unstable stuffed maps with one and two boundaries respectively, one can use a recursive formula to compute all higher genus and boundary generating functions \cite{blobbed1}. In particular for unstable stuffed maps, blobbed topological recursion reduces to the usual topological recursion. If more than the product of two traces occurs in the potential, then blobbed topological recursion is required. These computations are all done in Zhukovsky space.  
		
		\subsection{Unstable stuffed maps of genus zero with two boundaries}
		We would like to use blobbed recursion to compute higher order correlation functions of Dirac ensembles. In the previous section we showed how to compute $W_{1}^{0}(x)$ with various examples. The next step is to find $W_{2}^{0}(x_{1},x_{2})$, which for many important models,  in Zhukovsky space, is universal. For example this is true in single trace Hermitian matrix models \cite{Eynard2018} and in the multi-matrix model seen in \cite{two matrix}. As far as the authors of this paper can tell, the form of this function has not been established for multi-trace matrix models, despite efforts in this direction given in \cite{AK}. For our bi-tracial models the proof of the universal form of $W_{2}^{0}(x_{1},x_{2})$ is the same as the proof in Section 3.2 of \cite{Eynard2018}. That is 
		\begin{equation}\label{W2}
			W_{2}^{0}(x(z_{1}),x(z_{2}))  x'(z_{1})x'(z_{2}) d{z_{1}}d{z_{2}} = \left( \frac{1}{(z_{1}-z_{2})^{2}} - \frac{x'(z_{1})x'(z_{2})}{(x(z_{1})-x(z_{2}))^{2}}\right) d{z_{1}}d{z_{2}}.
		\end{equation}
		Notice the appearance of the Bergman kernel.
		
		If one wishes to compute the second order mixed moments, one needs to compute the residue 
		
		\begin{equation*}
			\mathcal{T}_{\ell_{1},\ell_{2}}= \text{Res}_{x_{1}\rightarrow \infty}\text{Res}_{x_{2}\rightarrow \infty}x_{1}^{\ell_{1}}x_{2}^{\ell_{2}} W_{2}^{0}(x_{1},x_{2})dx_{1}dx_{2}.
		\end{equation*}
		To simplify this calculation one may apply the Zhukovsky transform in both variables and see that the second term on the right hand side of equation (\ref{W2}) contributes nothing to the residue. 
		
		In summary, we have established that for our bi-tracial matrix models $W_{2}^{0}(x_{1},x_{2})$ has the same universal form as the single trace matrix models mentioned above.
		
		\subsection{Single trace models hidden in Dirac ensembles}
		Now with $W_{1}^{0}(x)$ and $W_{2}^{0}(x_{1},x_{2})$  we may compute higher order correlation functions of our Dirac ensembles.  This is proven as Theorem 9.1 in \cite{AK}. However, we shall not explicitly compute them here and instead have a different goal in mind.
		
		As noted in the previous section, the coupling constants of the quartic, cubic, and hexic Dirac ensembles  can be tuned to be the same as their respective Hermitian matrix model counterparts for certain values. When this is the case, $W_{1}^{0}(x)$ is the same, and in particular $\alpha$ and $\gamma$ are as well. Thus, combining this fact with the universal form of $W_{2}^{0}(x_{1},x_{2})$, via topological recursion all higher order genus expansion terms will be identical. Hence, we have proven that these single trace models hide in the above mentioned Dirac ensembles, at least for certain values of the coupling constants.

		In \cite{Eynard LQG} and chapter 5 of \cite{Eynard2018} it is proven that single trace Hermitian matrix models have interesting behavior at certain critical points. For the quartic model this occurs at $t_{4} = -\frac{1}{12}$, the hexic at $t_{4} = -\frac{1}{9}$ and $t_{6} = \frac{1}{270}$, and the cubic at $t_{3} = -\frac{1}{2}\,3^{-3/4}$. These points can be found as the locations of cusps of the spectral curve. See chapter 5 of \cite{Eynard2018} for details. Fortunately in the above mentioned Dirac ensembles all these critical points can be recovered. Their importance will be discussed in the following section.

		\section{The double scaling limit and 2D quantum gravity}
		In this section we will first review an old connection between random matrix theory and two dimensional quantum gravity. We will then discuss how these exact same connections hold for  specific examples of Dirac ensembles of dimension one.

		\subsection{Large maps}
		As discussed in Section \ref{Stuffed maps}, formal matrix integrals count maps which are essentially polygonizations of  Riemann surfaces. Intuitively, as the number of  2-cells that make up a map  increases, the polygonization should give a better approximation of the underlying surface. Thus, our goal is to fine-tune the coupling constants of the model to some critical point that will cause the number of polygons that make up maps to go to infinity. 
		
		We emphasize that this is not a new idea. It was know on a heuristic level to physicists in the 80's and 90's \cite{LQG 1,LQG 2} for asymptotic quantities of random matrix models. Physicists predicted a connection to Liouville conformal field theory coupled to gravity, and it was indeed proven using the KPZ formula \cite{BPZ}, that many models from both theories have the same critical exponents. Correlation functions of certain conformal field theories have the symmetry of  representations of conformal groups. Such infinite representations in two dimensions were classified in Kac's table \cite{Kac's table}, and distinguished by two integers $(p,q)$. For each such integer pair of a so-called minimal model the generating functions must satisfy a partial differential equation.  It was proved in \cite{Eynard LQG} that formal single matrix models have an associated $(2m+1,2)$ minimal model whose generating functions in the double scaling limit satisfy the associated partial differential equation and whose critical exponents match the minimal model. For example, the $(3,2)$ minimal model is referred to as pure gravity. Its generating functions satisfy Painlev\'e I. Both the quartic and cubic matrix models are associated with this minimal model. For a more detailed description see Chapter 5 of \cite{Eynard2018}.
		
		This idea is particularly useful in the context of Dirac ensembles because it provides a direct link between path integrals over metrics of finite noncommutative geometries to  two dimensional quantum gravity coupled to conformal field theories. We will describe this idea here. Consider the genus expansion of the partition function for some Dirac ensemble \cite{Second paper}
		\begin{equation*}
			\log Z := \sum_{g=0}^{\infty}N^{2-2g}F_{g},
		\end{equation*}
		where
		\begin{equation*}
			F_{g}  =\sum_{v=1}^{\infty}t^{v}\sum_{\Sigma \in \mathbb{SM}_{k}^{g}(v)} \prod_{i,j=1}^{d}t_{i,j}^{n_{i,j}(\Sigma)}\frac{1}{|\text{Aut} (\Sigma)|}.
		\end{equation*}
		In general, $F_{g}$ is a  function of the coupling constants with some algebraic  or logarithmic singularities \cite{Eynard2018}. These $F_{g}$'s can be computed from the $W_{k}^{0}$'s found using topological recursion. For details see Section 3.4 of \cite{Eynard2018}. As discussed in the previous section, for specific values of coupling constants the quartic, cubic, and hexic Dirac ensembles are precisely single trace Hermitian matrix models. Furthermore, all higher order correlation functions   $W_{k}^{0}$ are the same in these ranges of coupling constants. Thus, all $F_{g}$ are as well.
		
		As the number of vertices gets very large, the behavior of the $F_{g}$'s is going to be dominated by their singularities closest to zero. Thus, the behavior of unstable stuffed maps with a very large number of vertices is going to be controlled by the behavior of their generating functions near where its derivative diverges.
		
		Now consider the quartic Dirac ensemble as an example. We can fine tune its coupling constants such that we recover the quartic Hermitian matrix model. As seen in \cite{Eynard2018}, the explicit forms  of the $F_{g}$'s of the quartic matrix model imply that near a critical point, the series has an expansion as the sum of regular and singular functions. With $g\not = 1$ at the critical point $t_{4} = 1/12$, the singular part of the expansion looks like   
		\begin{equation*}
			\text{sing}(F_{g}) = C_{g}\, (t_{4}-t_{c})^{5(1-g)/2}
		\end{equation*}
		for some constant $C_{g}$. When $g =1$,
		\begin{equation*}
			\text{sing}(F_{1}) = C_{1} \,\log(t_{4}-t_{c}).
		\end{equation*}
		If one defines the new series 
		\begin{equation*}
			u(y) = \sum_{g= 0}^{\infty} C_{g}y^{5(1-g)/2},
		\end{equation*}
		then  $u''(y)$ satisfies the Painlev\'e I equation 
		\begin{equation*}
			y = (u''(y))^{2} - \frac{1}{3} u^{(4)}(y).
		\end{equation*}
		
		Note that, even in the double scaling limit, the generating functions of large maps can be computed using topological recursion \cite{Eynard2018}. Furthermore, for general single trace matrix models, we can construct the following formal series:
		\begin{equation*}
			\sum_{g\geq 0}N^{2-2g} \tilde{F}_{g},
		\end{equation*}
		where $\tilde{F}_{g}$ are the leading terms in the asymptotic expansions of $F_{g}$. This series is a formal Tau-function of some
		reduction of the KdV hierarchy, and can be obtained from Liouville conformal field theory when coupled with gravity.  For more details see Section 5.4 of \cite{Eynard2018}. Thus, if we can fine-tune a Dirac ensemble to the critical points of a single trace matrix model, the same result follows. In particular, we know that the critical point of the quartic Hermitian matrix model is $t_4 = -1/12$, and considering the analysis done in Section \ref{quartic analysis} we have the following phase diagram for the quartic Dirac ensemble.
		\begin{figure}[H]
			\centering
			\includegraphics[width=0.5\textwidth]{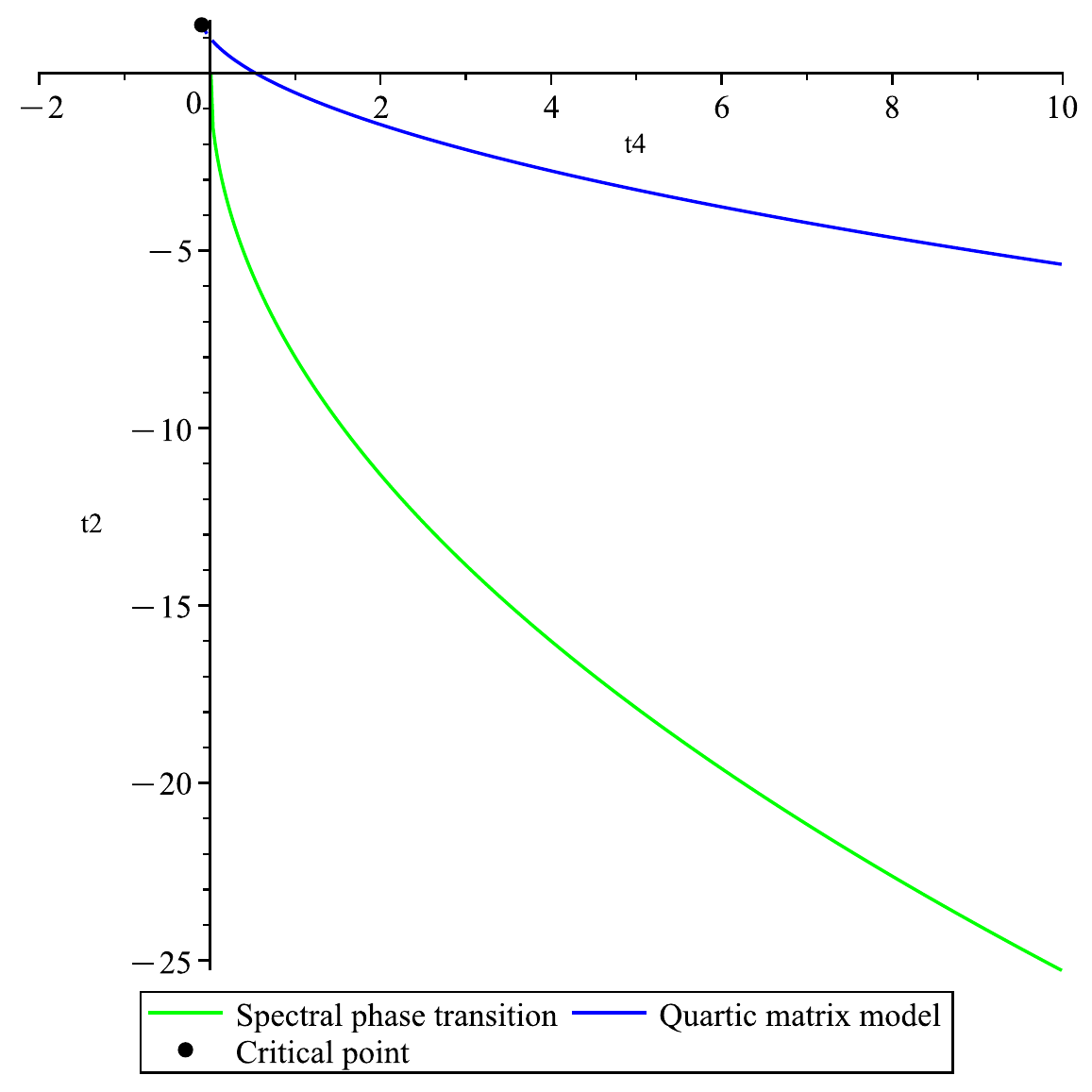}
			\caption{The phase diagram of the quartic Dirac ensemble.}
			\label{fig:quartic diagram}
		\end{figure}
		In the above diagram we have the curve where the spectral phase transition happens. This is precisely when the constant $t_2$ is chosen in terms of $t_{4}$ as
		\begin{equation*}
			t_2 =-8 \sqrt{t_4}.
		\end{equation*}
		This can be found by finding when $\rho(0) = 0$, just as in \cite{First paper}.
		
		One can also conclude that the quartic  matrix model appears when  
		\begin{equation*}
			t_2=-\frac{(1 + 12 t_4)^{3/2} - 4 - 144 t_4 + (36 t4 + 3) \sqrt{1 + 12 t_4}}{72 t_4}.
		\end{equation*}
		In the quartic matrix model the critical point occurs at $t_{4} = -\frac{1}{12}$. In the quartic Dirac ensemble this corresponds to the point $t_{4}= -\frac{1}{12}$ and $t_{2}=\frac{4}{3}$. Note that the spectral phase transition curve does not cross the quartic matrix model curve. This is because, in the quartic matrix model considered,
		
		\begin{equation*}
			\int_{\mathcal{H}_{N}}e^{-\frac{N}{2}\tr H^{2} -  \frac{c_{4}}{ 4}N\tr H^{4}}dH,
		\end{equation*}
		there is no spectral phase transition. A coupling constant in front of the Gaussian term is required for this phenomenon to occur.
		
		\begin{remark}
			\normalfont Ultimately we are solving the loop equations,  which are the same in this case regardless of whether the model is considered formal or convergent. Which interpretation we can choose is dependent upon which quadrants of Figure \ref{fig:quartic diagram} we are considering. In the first and fourth quadrants $t_{4}$ is negative in the potential, so the model can always be seen as convergent since the quartic term dominates the Gaussian one at the limits of the integral. In quadrants one and two the model can always be viewed as formal since the Gaussian integrals in the definition of a formal integral are always convergent. In the third quadrant, however, the model is neither formal nor convergent, as the integrals diverge, but so do all the Gaussian integrals in the formal definition.
		\end{remark}

\begin{figure}[H]
			\centering
			\includegraphics[width=0.5\textwidth]{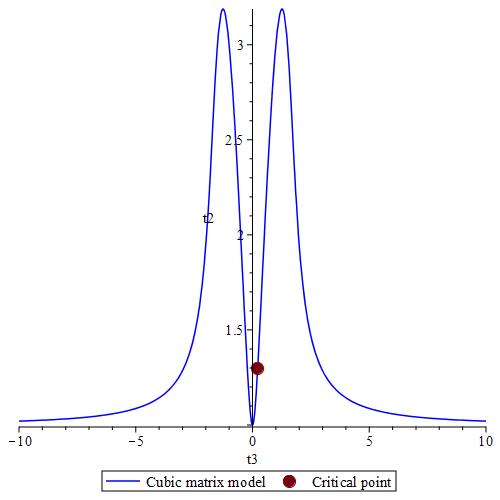}
			\caption{The phase diagram of the cubic Dirac ensemble.}
			\label{fig:cubic diagram}
		\end{figure}

		A similar analysis may be done for the cubic Dirac ensemble. However, no spectral phase transition occurs in this model. See  \ref{fig:cubic diagram}.
		The critical point is at $t_3 = -\frac{1}{2}\, 3^{-3/4}$ and 
		\begin{align*}
			t_2 &= {\frac {\sqrt {3}}{216}}+{\frac{1}{3}}
			+ {\frac {3168\,\sqrt {3}+ \left( 411915\,\sqrt {3}+418608+648\,\sqrt {
						239311+18512\,\sqrt {3}} \right) ^{{\frac{2}{3}}}+5043}{216\,\sqrt [3]
					{411915\,\sqrt {3}+418608+648\,\sqrt {239311+18512\,\sqrt {3}}}}}\\
			& \approx 1.297.
		\end{align*}

		In general it was shown in \cite{LQG 1} that a matrix model of the form 
		\begin{equation*}
			\int_{\mathcal{H}_{N}}e^{-\frac{N}{2}\tr H^{2} - \sum_{j\geq}^{d} \frac{c_{j}}{ j}N\tr H^{j}}dH
		\end{equation*}
		is associated with a  $(2m+1,2)$ minimal model in the double scaling limit. In particular, we can deduce then that the quartic, cubic, and hexic Dirac ensembles of type $(1,0)$ are associated with the $(3,2)$, $(3,2)$, and $(5,2)$ minimal models. In general for single matrix models, if the potential of the model is degree $d$ and it is odd then it is associated with the $(d,2)$ minimal model. If the degree is even, then it is associated with the $(d-1,2)$  minimal model. We expect the same relationship holds for Dirac ensembles of the above mentioned form.

		\section{Conclusion and Outlook}
		
		In this paper we analyze the quartic, cubic, and hexic Dirac ensembles of type $(1,0)$ as formal matrix integrals. Their resolvent functions $W_{1}^{0}$ and limiting eigenvalue distributions are found explicitly. The cylinder amplitude $W_{2}^{0}$ is discussed to have a universal form. Thus, via the process of blobbed topological recursion one may compute all higher genus and boundary correlation functions. During this analysis, it is found that by fine-tuning the coupling constants of these models one can recover critical phenomena seen in certain Hermitian matrix models. In particular in the  double scaling limit we find the critical exponents associated with minimal models from conformal field theories. Additionally, for these models the genus expansion terms of the log of the partition function satisfy the same differential equations as the partition functions of the corresponding minimal model in the double scaling limit. We hope to prove rigorously in future work that most Dirac ensembles of type $(1,0)$ and $(0,1)$ have a corresponding minimal model in the double scaling limit. Further, it would be interesting to construct multicritical matrix models \cite{multicrit 1,multicrit 2,multicrit 3} from Dirac ensembles. 
		
		Essentially, we have recovered conformal field theories coupled to gravity from toy models of quantum gravity on noncommutative spaces. We would like to extend this connection to Dirac operators of higher dimension. This seems likely considering that similar connections have been made for multi-matrix models \cite{Kazakov multi}. Without any known analytic techniques to study matrix models  seen in higher dimensional Dirac ensembles we can only aim for numerical evidence. It may be possible to deduce the critical values of Dirac ensembles using the bootstraps technique \cite{HKP}. In random matrix models the range of coupling constants on which the model is defined  determines the critical values. For example the quartic matrix model has solutions when $-1/12<t_4$ in the large $N$  limit. Thus, bootstrapping seems like a likely candidate to use to find these critical points. However, determining these critical exponents might be better suited to Monte Carlo simulations. This was explored in some sense in \cite{glaser}. Note that the finite size of $N$ might strongly affect these values. Another possibility is through functional renormalization group techniques \cite{Sanchez2}. We hope to explore these ideas in future works.
		
		We would like to thank the referees of this paper for their very thorough feedback and suggestions that have greatly improved this paper from its original draft.

	\end{document}